\documentclass[11pt]{article}

\usepackage{cite}
\usepackage{amssymb}
\usepackage{amsmath}
\usepackage{bm}

\begin{document}

\title{Weak Values: Approach through the Clifford and Moyal Algebras.}
\author{B. J. Hiley\footnote{E-mail address b.hiley@bbk.ac.uk.}.}
\date{TPRU, Birkbeck, University of London, Malet Street, \\London WC1E 7HX.}
\maketitle

\begin{abstract}
In this paper we calculate various transition probability amplitudes, TPAs, known as `weak values'  for the Schr\"{o}dinger and Pauli particles. It is shown that these values are related to the Bohm momentum, the Bohm energy and the quantum potential in each case.  The results for the Schr\"{o}dinger particle are obtained in three ways, the standard approach, the Clifford algebra approach of Hiley and Callaghan, and the Moyal approach.  To obtain the results for the Pauli particle, we combine the Clifford and Moyal algebras into one structure. The consequences of these results are discussed.

\end{abstract}

\section{Introduction.}
Let us begin by considering the mean value of some Hermitean operator $\hat A$ in the state $|\psi\rangle$. By introducing a complete orthonormal set $|\phi_i\rangle$, we can write
\begin{eqnarray}
\langle \psi|\hat{A}|\psi\rangle =\sum_i\langle\psi|\phi_i\rangle\langle\phi_i|\hat{A}|\psi\rangle=\sum_i \rho_\psi(\phi_i) \frac{\langle\phi_i|\hat{A}|\psi\rangle}{\langle\phi_i|\psi\rangle}	\label{1}
\end{eqnarray}
In recent literature, the quantity $ \langle\phi_i|\hat{A}|\psi\rangle/\langle\phi_i|\psi\rangle$  has been called a `weak value' because its magnitude can be found in a so called `weak measurement' if certain criterion are met \cite{yadalv88} \cite{idpses89}.   However in the context of this paper the concept of `weakness'  has no meaning.  It is merely a transition probability amplitude, [TPA], which, in general, will be a complex number. Nonetheless we will, following convention, continue to call it a weak value.

The choice of  $|\phi_i\rangle$ is known as `post selection'.  Suppose we post select $|a_i\rangle$, an eigenket of $\hat{A}$, we then find
\begin{eqnarray}
\frac{\langle a_i|\hat{A}
|\psi\rangle}{\langle a_i|\psi\rangle}=a_i, \label{2}
\end{eqnarray}
so that  equation (\ref{1}) becomes simply
\begin{eqnarray*}
\langle \psi|\hat{A}|\psi\rangle =\sum_i \rho_\psi(a_i) a_i
\end{eqnarray*}
This is an extremely well known result showing that in this particular case the TPA (\ref{2}) is simply an eigenvalue $a_i$.  Here it is meaningful to call (\ref{2})  a `value' as it has that meaning, but again there is nothing `weak' about it.

Let us now consider a more interesting case where we post select $|b_i\rangle$, an eigenket of an operator $\hat{B}$ which does not commute with $\hat {A}$.  Now form
\begin{eqnarray*}
\langle \psi|\hat{A}|\psi\rangle =\sum_{i} \rho_\psi(b_i)\frac{\langle b_i|\hat{A}|\psi\rangle}{\langle b_i|\psi\rangle}.
\end{eqnarray*}
The last term in this equation can be written as
\begin{eqnarray*}
\frac{\langle b_i|\hat{A}|\psi\rangle}{\langle b_i|\psi\rangle}=\sum_{j}a_j\frac{\langle b_i|a_j\rangle\langle a_j|\psi\rangle}{\langle b_i|\psi\rangle}
\end{eqnarray*}
Thus we see that in this case, the meaning of the TPA is not a simple `value' of anything.  Here it is about a relationship between values.

A more interesting example is when we replace $\hat{A}$ and $\hat{B}$ by a position operator $\hat{X}$ and the momentum operator $\hat{P}$.  In this case we have
\begin{eqnarray*}
\langle\psi(t)|\hat{P}|\psi(t)\rangle=\int\rho(x,t)\frac{\langle x|\hat{P}|\psi(t)\rangle}{\langle x|\psi(t)\rangle}dx.
\end{eqnarray*}
To take this further, we find 
\begin{eqnarray*}
\langle x|\hat{P}|\psi(t)\rangle=\int\langle x|\hat{P}|x'\rangle\langle x'|\psi(t)\rangle dx'
\end{eqnarray*}
But $\langle x|\hat{P}|x'\rangle=-i\delta(x-x')\nabla_{x'}$,\footnote{We have put $\hbar=1$ through out this paper.} so that
\begin{eqnarray*}
\langle x|\hat{P}|\psi(t)\rangle=-i\nabla_x\psi(x,t).
\end{eqnarray*}
Let us now write $\psi(x,t)=R(x,t)e^{iS(x,t)}$ and find
\begin{eqnarray}
\frac{\langle x|\hat{P}|\psi(t)\rangle}{\langle x|\psi(t)\rangle}=\nabla_x S(x,t)-i\nabla_x \rho(x,t)/2\rho(x,t).		\label{3}
\end{eqnarray}
Here $\rho(x,t)$ is the usual expression for the probability, $\rho(x,t)=|\psi(x,t)|^2$.  Thus we see this weak value is complex, the real part simply being the Bohm momentum.  

This fact has already been pointed out by Leavens \cite {crl05}, but he makes no comment on the imaginary part.
In fact the imaginary part is identical to the `osmotic velocity' \footnote{The notion of an `osmotic velocity' arises whenever there is a probability gradient no matter what is the cause of this gradient. Here we offer no underlying physical reason for the existence of this gradient in quantum processes.} introduced by Bohm and Hiley \cite{dbbh89} in their exploration of the relation of the Bohm model to a diffusion model first introduced by Nelson \cite{en85}.

Before going on to discuss the possible relationship between a diffusion model of quantum phenomena to the standard approach, I want to draw attention to a similar result obtained by Wiseman \cite{hmw07} in a slightly different manner.  

He starts by recalling that, in the Heisenberg picture, the velocity can be obtained from the equation
\begin{eqnarray*}
i\frac{d\hat{X}}{dt}=[\hat{X},\hat{H}].
\end{eqnarray*}
By using the definition of a weak value, we find
\begin{eqnarray}
v(x,t)=Re\frac{\langle x|i[\hat{H}, \hat{X}]|\psi(t)\rangle}{\langle x|\psi(t)\rangle}.   \label{4}
\end{eqnarray}
Then if we put $\hat{H}=\hat{P}^2/2m +V(\hat{X})$ and $\psi(x,t)=R(x,t)e^{iS(x,t)}$, we again get an expression of the Bohm velocity $\nabla_x S(x,t)/m$.  Wiseman  goes on to discuss the physical and philosophical implications that stem from this result.  Here we will only concentrate on the mathematical structure lying behind these weak values, extending them to the case of a non-relativistic particle with spin.

\section{ Relationship of these results to Nelson's discussion}

\subsection{The left-right derivatives, the Bohm Momentum and the Moyal Marginal Momentum.}

Let us explore the relation to Nelson's work \cite{en85}. He writes equation (\ref{3}) as a velocity in the form  
\begin{eqnarray}
v(x,t)=\frac{1}{m}\nabla_x S(x,t)-iD\frac{\nabla_x \rho(x,t)}{\rho(x,t)}.			\label{5}
\end{eqnarray}
If we write $D=1/2m$ in equation (\ref{5}), we immediately see that the last term has exactly the same form as the osmotic velocity introduced by Nelson \cite{en66} and also later by Bohm and Hiley \cite{dbbh89}.  However we have derived the result directly from a standard weak value and have no need to speculate about underlying diffusive motion. 

To continue let us write  the Bohm velocity, $v_B(x,t)=\nabla_x S(x,t)/m$, in the form
\begin{eqnarray}
v_{B}(x,t)=\frac{1}{2m}\left[ \frac{\langle \psi(t)|\hat{P}|x\rangle}{\langle \psi(t)|x\rangle} +\frac{\langle x|\hat{P}|\psi(t)\rangle}{\langle x|\psi(t)\rangle}     \right]
	\label{7}
\end{eqnarray}
while the osmotic velocity, $v_O$, is
\begin{eqnarray}
v_{O}(x,t)=\frac{1}{2m}\left[\frac{\langle \psi(t)|\hat{P}|x\rangle}{\langle \psi(t)|x\rangle} -\frac{\langle x|\hat{P}|\psi(t)\rangle}{\langle x|\psi(t)\rangle}      \right]
	\label{8}
\end{eqnarray}
Now let us compare these results with those obtained in Bohm and Hiley \cite{dbbh89}. They write 
\begin{eqnarray}
v_B(x,t)= \left[b(x,t)+b_*(x,t)\right]/2 \quad\mbox{and}\quad
v_O(x,t)=b(x,t)-b_*(x,t).   \label{6}
\end{eqnarray}
Here $b(x,t)$ is the mean `forward' velocity, which is based on Nelson's mean forward derivative \cite{en85},
 \begin{eqnarray*}
  Dx(t)=\lim_{\Delta t\rightarrow 0^+}E_i\frac{x(t+\Delta t)-x(t)}{\Delta t}
 \end{eqnarray*}
 and $b_*(x,t)$ is the mean `backward' velocity, which in turn is based on Nelson's mean backward derivative 
 \begin{eqnarray*}
  D_*x(t)=\lim_{\Delta t\rightarrow 0^+}E_i\frac{x(t)-x(t-\Delta t)}{\Delta t}
 \end{eqnarray*}
 The need to introduce two derivatives in a diffusive or stochastic process arises because $x(t)$ is not differentiable.  It is then necessary to distinguish between the probability that a particle initially at position $x-\Delta x$ will arrive at $x$ at time $t$ in the time interval $\Delta t$ and the probability that a particle at $x +\Delta x$ would have been found at $x$ at an earlier time $t-\Delta t$.
 
 Comparing the above results, we find that, in the quantum formalism, we may identify these velocities with a weak value in the following way,
\begin{eqnarray*}
b(x,t)=\frac{1}{m}\frac{\langle \psi(t)|\hat{P}|x\rangle}{\langle \psi(t)|x\rangle}\quad\mbox{and}\quad b_*(x,t)=\frac{1}{m}\frac{\langle x|\hat{P}|\psi(t)\rangle}{\langle x|\psi(t)\rangle}
\end{eqnarray*}
This difference, rather than arising from a diffusive motion, arises from formal noncommutativity.  Symbolically we can say that right action $\overrightarrow P|\psi(t)\rangle$ is different from left action $\langle\psi(t)|\overleftarrow P$.  However as can be seen from equations (\ref{7}) and (\ref{8}),  the physically meaningful quantities arise from sums and differences of weak values, so we expect terms like 
\begin{eqnarray}
\frac{\langle \psi(t)|\hat{P}|x\rangle}{\langle \psi(t)|x\rangle}\pm \frac{\langle x|\hat{P}|\psi(t)\rangle}{\langle x|\psi(t)\rangle}=\frac{\psi^*(x)\overleftarrow\nabla_x\psi(x)\pm\psi^*(x)\overrightarrow\nabla_x\psi(x)}{\rho(x)}	\label{9}
\end{eqnarray}
to play a significant role.

\subsection{The Moyal Algebra.}

Let us now see how these weak can be treated by using the Moyal algebra \cite{jem49}, \cite{rgl86} \cite{cz02}.  Recall this algebra does not use operators but instead uses ordinary functions of $x$ and $p$.  Here the distinction between left and right action arises as  a consequence of the noncommutativity of the Moyal $\star$-product.  We recall that the $\star$-product, \cite{rgl86}, \cite{cz02}, is defined by 
\begin{eqnarray*}
\star=\exp\left[\frac{i}{2}\left(\overleftarrow{\partial_x}\overrightarrow{\partial_p}-\overleftarrow{\partial_p}\overrightarrow{\partial_x}\right)\right]
\end{eqnarray*}
Then we find\footnote{We will suppress the time dependence for succintness} 
\begin{eqnarray*}
p\star f(x,p)=\left(p-\frac{i}{2}\overrightarrow{\partial_x}\right)f(x,p)
\end{eqnarray*}
and the dual 
\begin{eqnarray*}
f(x,p)\star p=f(x,p)\left(p+\frac{i}{2}\overleftarrow{\partial_x}\right)
\end{eqnarray*}
Here we see clearly the need for a left and right derivatives.  The sums and differences of the type shown in equation (\ref{9}) suggests we form the Baker\footnote{This is just the Jordon product.} and the Moyal brackets which are defined as
\begin{eqnarray*}
[p,f(x,p)]_{BB}=\frac{f\star p+p\star f}{2}=pf(x,p)
\end{eqnarray*}
and
\begin{eqnarray*}
[p,f(x,p)]_{MB}=\frac{f\star p-p\star f}{i}=\nabla_x f(x,p).
\end{eqnarray*}
To this point we have assumed $f(x,p)$ is some general phase space function.  Let us now choose $f(x,p)$ to be the  Wigner-Moyal distribution, given by
\begin{eqnarray}
f(x,p)=2\pi\int \psi^*(x-y/2)\psi(x+y/2)e^{ip.y}dy
\label{WM}
\end{eqnarray}
Alternatively, as Moyal shows in \cite{jem49}, we may also express this relationship in terms of a momentum representation.  This is just the Fourier transform of equation (\ref{WM}).  Thus
\begin{eqnarray*}
f(x,p)=\left(\frac{1}{2\pi}\right)\int \phi^*(p+\theta/2)\phi(p-\theta/2)e^{-ix.\theta}d\theta.
\end{eqnarray*}
This equation can also be written as
\begin{eqnarray}
f(x,p)=2\pi\int\int\phi^*(p_1)\phi(p_2)\delta\left(p-\frac{p_2+p_1}{2}\right)e^{ix.(p_2-p_1)} dp_1dp_2		\label{WMP}
\end{eqnarray}
To make contact with the results of subsection (2.1), we simply project to marginals by either integrating over the momentum, $p$, or over the position $x$, to find
\begin{eqnarray*}
\int f(x,p)dp=|\psi(x)|^2=\rho(x)\quad\mbox{and}\quad\int f(x,p)dx=|\phi(p)|^2=\bar{\rho}(p),
\end{eqnarray*}
which are just the standard probability densities. The marginal momentum, $\overline{\overline p}(x)$, introduced by Moyal \cite{jem49}, is given by
\begin{eqnarray*}
\rho(x)\overline{\overline{p}}(x)=\int pf(x,p)dp.
\end{eqnarray*}
Using equation (\ref{WMP}) we find,
\begin{eqnarray}
\rho(x)\overline{\overline{p^n}}(x)=2\pi\int\int p^n\phi^*(p_1)\phi(p_2)\delta\left(p-\frac{p_2+p_1}{2}\right)e^{ix.(p_2-p_1)} dp_1dp_2\nonumber\\
=\left(\frac{1}{2i}\right)^n\left[\left(\frac{\partial}{\partial x_1}-\frac{\partial}{\partial x_2}\right)^n\psi(x_1)\psi^{\star}(x_2)\right]_{x_1=x_2=x}  \label{6}
\end{eqnarray}
In the case when $n=1$ and $\psi=Re^{iS}$ it can easily be shown that
\begin{eqnarray*}
\overline{\overline{p}}(x)=\nabla_x S(x).
\end{eqnarray*}
which is, of course, the Bohm momentum.

The corresponding integration of the Moyal bracket gives
\begin{eqnarray*}
\frac{\int [p,f(x,p)]_{MB}dp}{2\rho(x)}=\frac{\nabla_x \left(\int f(x,p)dp\right)}{2\rho(x)}=\frac{\nabla_x \rho(x)}{2\rho(x)} 
\end{eqnarray*}
showing clearly that what has been called the `osmotic' term arises from the failure of the Moyal bracket to vanish, i.e., the noncommutativity of the Moyal product.  

Thus we have derived the same results as obtained from the left-right derivatives $\overrightarrow{P}|\psi\rangle$ and $\langle\psi|\overleftarrow{P}$.  In this particular case the Baker and Moyal brackets are simply a way of separating the real and the imaginary parts of weak values. 
 It should also be noticed that the Bohm model and the Moyal approach produce  identical results as has already been shown by Hiley \cite{bh10}.

\subsection{Consequences for the Kinetic Energy.}

To complete this part of the discussion and show a further relation of weak values to the Bohm model, let us examine what happens to the kinetic energy.  Consider the second order derivative $p^2$, and form
\begin{eqnarray*}
p^2\star f(x,p)=\left[p^2-i p\;\overrightarrow{\nabla_x}-\frac{1}{4}\;\overrightarrow{\nabla_x}^2\right]f(x,p).
\end{eqnarray*}
together with 
\begin{eqnarray*}
f(x,p)\star p^2=f(x,p)\left[p^2+ip\;\overleftarrow{\nabla_x}-\frac{1}{4}\;\overleftarrow{\nabla_x}^2\right]
\end{eqnarray*}
Then forming the Baker bracket we find
\begin{eqnarray*}
[p^2,f(x,p)]_{BB}=p^2f(x,p)-\frac{1}{4}\nabla^2_x f(x,p).
\end{eqnarray*}
If we again integrate over $p$ to obtain the marginal kinetic energy\footnote{For simplicity we assume a mass of the particle to be 1/2.},
\begin{eqnarray*}
\int [p^2,f(x,p)]_{BB}dp=\rho(x)\overline{\overline{p^2}}(x)-\frac{1}{4}\nabla_x^2\rho(x).
\end{eqnarray*}
Using (\ref{6}) for $n=2$, we find
\begin{eqnarray*}
\overline{\overline{p^2}}(x)=(\nabla_x S(x))^2+\frac{1}{4}
\frac{\nabla_x^2\rho(x)}{\rho(x)}-\frac{\nabla_x^2R(x)}{R(x)},
\end{eqnarray*}
so that 
\begin{eqnarray*}
\int [p^2,f(x,p)]_{BB}dp/\rho(x)=(\nabla_xS(x))^2-\nabla_x^2R(x)/R(x).
\end{eqnarray*}
The Bohm kinetic energy for a particle of mass 1/2 is $KE_B =(\nabla_x S)^2$, while the second term is the quantum potential $Q=-\nabla_x^2R/R$.  Thus we see that the integral over $p$ of the Baker bracket, produces both the Bohm kinetic energy and the quantum potential for the free particle again showing another way to establish a relation between the weak values and the Bohm variables.

This leaves us with the Moyal bracket which we write down for completeness is 
\begin{eqnarray*}
\int [p^2,F(x,p)]_{MB}dp/\rho(x)=\nabla_x^2S(x)+\left(\frac{\nabla_x \rho(x)}{\rho(x)}\right)\nabla_xS(x).
\end{eqnarray*}

It should also be noted that these results are identical with those obtained from 
$
\left(\langle\psi(t) |\hat P^2|x\rangle\pm\langle x|\hat {P^2}|\psi(t)\rangle\right)/2$,		
 showing that the Baker and Moyal brackets give exactly the same results as sums and differences of weak values.
 
We end this section by making two remarks.

(1) Exactly the same results can be obtained using cross Wigner functions as has been recently shown by de Gosson and de Gosson \cite{mdegsdeg11}.

(2) All these results show that the TPA or `weak' value approach, the Bohm approach and the Moyal approach produce identical results.

\section{Extension to Spin-half Particles.}

Let us now move on to consider weak values for a non-relativistic particle with spin-half, namely the Pauli particle.  In this case there are two methods that we could use.  The straight forward but somewhat inelegant conventional method using matrices or the approach using Clifford algebras, the greater details of which were presented in Hiley and Callaghan \cite{bhrc11A}.  However this approach introduces the concept of a generalised Dirac operator \cite{jgmm} which was imported from outside the algebraic structure.  Furthermore  it does not show that these derivatives introduce a generalised symplectic structure.

  What we will do now is to show how the generalised Dirac derivative can emerge from a larger algebra formed by combining the Clifford and Moyal algebras.  Thus both the orthogonal and symplectic (Moyal) structures are united into a single algebraic structure
\cite{ac90}. 

  In our previous work  \cite{bhrc10} \cite{bhrc11A}, we used the elements $\Phi_L(x)$ and its conjugate, $\widetilde{\Phi}_L(x)$, formed by Clifford reversion, to construct an entity which we have called the Clifford density element $\rho_c(x)=\Phi_L(x)\widetilde{\Phi}_L(x)$.  In this paper we will generalise this object to include wider class of entities by writing
\begin{eqnarray*}
\rho_{\nu,\mu}(x_1,x_2)=\Phi_{L_{\nu}}(x_1)\widetilde{\Phi}_L{_{\mu}}(x_2)
\end{eqnarray*}
so that $\rho_c(x)=\rho_{\nu,\nu}(x_1=x_2=x)$\footnote{For the most general case we would write $\rho_{\Phi,\Xi}=\Phi_L(x_1)\Xi_R(x_2)$, but we will not pursue this generalisation here.}.
Then we construct a generalisation of equation (\ref {WM})
\begin{eqnarray*}
F(x,p)=2\pi\int Tr[\Phi_L(x_1)\widetilde{\Phi}_L(x_2)] e^{-ip.y}dy.
\end{eqnarray*}
where $x=(x_2 +x_1)/2$ and $y=(x_2-x_1)$ and Tr is a trace.  It should be noted that we have chosen the notation $F(x,p)$ as it is a generalisation of the Moyal density $f(x,p)$ for spin-less particles as we will explain below. 

In the previous papers \cite{bhrc10}, \cite{bhrc11A}, we formed an element of a minimal left ideal in the Pauli Clifford algebra ${\cal C}_{3,0}$ by choosing a particular primitive idempotent $\epsilon=(1+e_3)/2$ and then constructed elements of the form
\begin{eqnarray}
\Phi_L(x)=\phi_L(x)\epsilon=[g_0(x) +g_1(x)e_{23}+g_2(x)e_{13}+g_3(x)e_{12}]\epsilon.		\label{Phi}
\end{eqnarray}
Here $e_{ij}=e_i\circ e_j$ where the $e_k$ are the generating elements of ${\cal{C}}_{3,0}$, the Pauli Clifford algebra\footnote{The generators of the Clifford algebra, $e_k$, combine using  the Clifford product,\\ $[e_i\circ e_j+e_j\circ e_i]=2\delta_{ij}$}.  However, for our purposes here, there is a more convenient, but equivalent way of writing equation (\ref{Phi}), namely,
\begin{eqnarray*}
\Phi_L(x)=[(g_0-g_3e_{123})+(g_2-g_1e_{123})e_1]\epsilon,
\end{eqnarray*}
so that if we use
\begin{eqnarray}
2g_0&=&\phi^*_1+\phi_1\quad \hspace{0.5cm}2g_2=\phi^*_2+\phi_2\nonumber\\
2e_{123}g_3&=&\phi^*_1-\phi_1\quad 2e_{123}g_1=\phi^*_2-\phi_2,
\label{g-phi}
\end{eqnarray}
$\Phi_L$ becomes simply
\begin{eqnarray*}
\Phi_L(x)=[\psi_1(x)+\psi_2(x)e_1]\epsilon.
\end{eqnarray*}

 {\noindent}Actually we can make things somewhat easier if we  express the result in terms of the $p$-representation.  Then
\begin{eqnarray}
F(x,p)=\left(\frac{1}{2\pi}\right)\int Tr[\Xi_L(p_1)\widetilde{\Xi}_L(p_2)] e^{-ix.\pi}d\pi.		\label{14}
\end{eqnarray}
where $p=(p_2+p_1)/2$ and $\pi=p_2-p_1$.  In this case we write
\begin{eqnarray}
\Xi_L(p)=\xi(p)\epsilon=[\phi_1(p)+\phi_2(p)]\epsilon		\label{15}
\end{eqnarray}
where, of course, the $\phi(p)$s are Fourier transforms of the $\psi(x)$s,

To find the Bohm momentum in the Moyal algebra, we use 
\begin{eqnarray*}
F(x,p)=\left(\frac{1}{2\pi}\right)\int Tr[\xi_L(p_2)\epsilon\widetilde{\xi}_L(p_1)]e^{-ix.\pi}d\pi
\end{eqnarray*}
with $\epsilon=(1+e_3)/2$.
Then using equation (\ref{15}), together with its conjugate
\begin{eqnarray*}
\widetilde{\Xi}_L(p)=\epsilon[\phi_1^*(p)+\phi_2^*(p)],
\end{eqnarray*}
we find
\begin{eqnarray}
F(x,p)=\int\int M(p_2,p_1)\delta\left(p-\frac{p_2-p_1}{2}\right)e^{ix.(p_2-p_1)}dp_1dp_2,	\label{22}
\end{eqnarray}
where
\begin{eqnarray}
M(p_2,p_1)=[\phi_1(p_2)\phi_1^*(p_1)+\phi_2(p_2)\phi_2^*(p_1)].		\label{M21}
\end{eqnarray}

Having obtained an expression for $F(x,p)$,  let us now consider
\begin{eqnarray*}
 p\star F(x,p)= p F(x,p) -\frac{i}{2}\nabla_xF(x,p)
\end{eqnarray*}
and
\begin{eqnarray*}
F(x,p)\star  p= p F(x,p) +\frac{i}{2}\nabla_xF(x,p)
\end{eqnarray*}
where we have written $p=\sum_{j=1}^3p_je_j$.   The Baker bracket and the Moyal bracket are then
\begin{eqnarray*}
[ p,F]_{BB}= pF(x,p),\quad\mbox{and}\quad [ p,F]_{MB}=\nabla_xF(x,p).
\end{eqnarray*}

Finally to obtain the expression for the Bohm momentum, we take the marginal
\begin{eqnarray*}
\rho(x) P_B(x)=\int [ p,F]_{BB}d p=\int  pF(x,p)d p.
\end{eqnarray*}

Now we need to evaluate the last integral in this equation.  To do this we note
\begin{eqnarray*}
\int pF(x,p)dp=\int\int\int pM(p_2,p_1)\delta\left(p-\frac{p_2-p_1}{2}\right)e^{ix.(p_2-p_1)}dp_1dp_2dp.
\end{eqnarray*}
The RHS of this equation is the same form as the second term in equation (\ref{6}) except that it contains the sum of two pairs of $\phi(p_2)\phi^*(p_1)$.  Each pair can be evaluated separately, so that finally we obtain
\begin{eqnarray}
\rho(x)P_B(x)=\int [p,F]_{BB}dp=\rho_1(x)\partial_x S_1(x)+\rho_2(x)\partial_x S_2(x),	\label{BBS}
\end{eqnarray}
where we have, as usual, used $\psi=Re^{iS}$.  This is exactly the expression for Bohm momentum presented in Bohm and Hiley \cite{dbbh93}.   As a final confirmation we need to show this corresponds to the expression obtained by Bohm, Schiller and Tiomno \cite{bst}. We do this by first writing the spinor wave function, $\Psi(x)$, in the form of the Caley-Klein parameters, viz,
\begin{eqnarray}
\Psi(x)= R(x)e^{iS(x)/2}   \begin{pmatrix} 
      \cos\theta(x)/2e^{i\phi(x)/2} \\
      i\sin\theta(x)/2e^{-i\phi(x)/2} \\
   \end{pmatrix}  	\label{ck}
\end{eqnarray}
then it is straight forward to show that the expression (\ref{BBS}) reduces to 
\begin{eqnarray*}
P_B=\nabla_x S/2 +\cos\theta\;\nabla_x \phi/2,
\end{eqnarray*}
which is just the expression found in Bohm, Schiller and Tiomno \cite{bst}

\subsection{The Bohm Kinetic Energy from the Moyal Algebra}

Let us complete this section by using the Moyal algebra to find the Bohm kinetic energy for the Pauli particle and an expression for its quantum potential.  The method is straight forward. Using
$
p^2=(\sum p_je_j)^2=p_1^2+p_2^2+p_3^2.
$
we find 
\begin{eqnarray*}
p^2\star F(x,p)=\left[p^2-i p\;\overrightarrow{\nabla_x}-\frac{1}{4}\;\overrightarrow{\nabla_x}^2\right]F(x,p).
\end{eqnarray*}
While  
\begin{eqnarray*}
F(x,p)\star p^2=F(x,p)\left[p^2+i p\;\overleftarrow{\nabla_x}-\frac{1}{4}\;\overleftarrow{\nabla_x}^2\right].
\end{eqnarray*}
Then forming the Baker bracket we find
\begin{eqnarray*}
[p^2,F(x,p)]_{BB}=p^2F(x,p)-\frac{1}{4}\nabla_x F(x,p).
\end{eqnarray*}

As we have seen in the previous subsection, $F(x,p)$ behaves as if it a sum of two $f(x,t)$, one corresponding to component $\psi_1$ of the spinor and the other corresponding to $\psi_2$.  Thus we find the marginal
\begin{eqnarray*}
\int [p^2,F(x,p)_{BB}]dp=\rho_1(x)\overline{\overline {p_1^2}}-\frac{1}{4}\nabla_x^2\rho_1(x)+\rho_2(x)\overline{\overline {p_2^2}}-\frac{1}{4}\nabla_x^2\rho_2(x)	
\end{eqnarray*}
which leads directly to the result
\begin{eqnarray}
KE_B=\rho_1(x)(\nabla_xS_1(x))^2 +\rho_2(\nabla_xS_2(x))^2\hspace{1cm}\nonumber\\-R_1(x)\nabla_x^2R_1(x)-R_2(x)\nabla_x^2R.	\label{EE}
\end{eqnarray}
Unfortunately this equation does not fall easily into the form $P^2_B/2m+Q$ so that it can be compared directly with the expression obtained in Bohm, Schiller and Tiomno \cite{bst}.  However using Cayley-Klein expression for the spinor $\Psi(x)$, the  RHS of equation (\ref{EE}) takes the form
\begin{eqnarray}
 P^2_B+Q=P^2_B-\frac{\nabla_x^2R}{R}+(\nabla_x\theta/2)^2+\sin^2\theta(\nabla_x\phi/2)^2		\label{QP}
\end{eqnarray}
This is exactly the form given in Bohm, Schiller and Tiomno \cite{bst} and also in Dewdney, Holland, Kyprianidis and Vigier \cite{DHKV} where there is a simple discussion of the whole approach.

This completes the demonstration of how the Moyal approach coupled with the orthogonal Clifford can be used to obtain weak values.

\section{Conclusions}

In this paper we have shown how a set of weak values can be given a simple meaning in terms of the Bohm  approach \cite{dbbh93} to quantum mechanics.  Not only have we shown that weak values can be given a meaning for the non-relativistic Schr\"{o}dinger particle, but they are also given meaning for a non-relativistic Pauli particle with spin-half.

In the case of the Schr\"{o}dinger particle, we have shown how these weak values can be obtained from the standard approach and through a approach using the Moyal algebra \cite{jem49}.  To extend the approach to include a spin-half particle, we have combined the Clifford algebra approach of Hiley and Callaghan \cite {bhrc10} \cite{bhrc11A} \cite{bhrc11B} with the Moyal algebra.  This gives a more general mathematical structure in which spin can naturally be included.  Space has prevented this method being applied to the relativistic Dirac particle.  These results, which are based on the work reported in \cite{bhrc11B} will be presented elsewhere.
 
 A union of the Clifford and Moyal algberas does not merely present a unified mathematical structure, but it is the very nature of this algebra that has significance.  The orthogonal Clifford was constructed by Clifford himself  \cite{wkc82}, not by considering quantum mechanics, but by arguing that it arises from movements or processes in Euclidean space-time. Clifford used terms like `versors' (literally `turners'), `rotors' and `motors' to try to emphasise his overarching philosophical idea;  an idea that claims everything starts with activity, with movement, i.e. from process --  and that the processes themselves condition the geometry of space-time \cite{dbbh93a} \cite{bh11}. 
 
In this sense the algebra is a geometric algebra and quantum phenomena are conditioned by, and help condition the geometry as has been stressed by Hestenes \cite{dh03} and by Doran and Lasenby \cite{cdal03}.  These authors considered only the orthogonal Clifford, but in this paper we have shown how the symplectic Clifford algebra \cite{ac90}, namely, the Moyal algebra, must also be included.
As remarked by Dubin, Hennings and Smith \cite{ddtbs00}, this algebraic structure brings out the deeper connection between the Moyal algebra and Weyl quantisation.  

The existence of this intimate relation between the extended algebra and quantum phenomena suggests that the latter are not something exotic, but deeply related to subtle topological structures within space-time itself \cite{lk88}.  In addition
this more general mathematical structure has the advantage that it shows a much closer relationship between quantum and classical phenomena than normally assumed. This point is discussed in detail in a recent paper by de Gosson and Hiley \cite{mdgbh11}.  A further merit of our approach is that the classical limit emerges naturally without any need to appeal to decoherence \cite{bh10}.

One of the key features of the approach using the orthogonal Clifford is the central role played by the Clifford density element $\rho_c=\Phi_L(x)\widetilde{\Phi}_L(x)$, where the tilde denotes Clifford reversion \cite{bhrc11A}.  This object is the algebraic analogue to the standard density matrix.  Its extension in order to evaluate weak values follows immediately by a generalisation which involves replacing $\widetilde{\Phi}_L(x)$ by an element of a different right ideal.  The further extension is needed to include the Moyal algebra.  This extension is 
\begin{eqnarray*}
\rho_M(x,p)=\int \Phi_L(x_1)\Xi_R(x_2)e^{ipy}dy,
\end{eqnarray*}
where $x=(x_2+x_1)/2$ and $y=(x_2-x_1)$. This is just the cross-Wigner function used in de Gosson and de Gosson \cite{mdegsdeg11}.  We see that we are constructing a non-local description of quantum phenomena.  The locality of the classical world then emerges as a limiting procedure which neglects the more subtle features of quantum processes \cite{dbbh93a}.

One final remark. The recent weak measurement experiments of Kocis, Braverman, Ravets, Stevens, Mirin, Shalm, and Steinberg \cite{ams11} and of Lundeen, Sutherland, Patel, Stewart and Bamber \cite{jsl11} opens up the possibility of actually measuring the Bohm momentum, the Bohm energy and hence the quantum potential.  This possibility moves the Bohm approach from a mere philosophical speculation into the realm of experimental physics.  However welcome as these results are, our discussion as to how these weak values arise strongly suggests that the so-called `Bohmian mechanics' is a pale reflection of the deeper algebraic structure underlying quantum processes.

\section{Acknowledgements}
I wish to thank Bob Callaghan, Maurice de Gosson, David Robson and Graham Yendell for many stimulating discussions.  



\bibliography{myfile}{}
\bibliographystyle{plain}

\end{document}